\documentclass[twocolumn,prl,amsmath,amssymb,showpacs,floatfix]{revtex4}
\usepackage{epsfig}
\usepackage{amstext,amssymb,amsmath}

\begin{document}

\title{Interconversion of dark soliton and Josephson vortex in a quasi-1D long Bose Josephson junction}
\author{V.M. Kaurov and A.B. Kuklov}
\affiliation
{Department of Engineering Science and Physics,
The College of Staten Island, CUNY, 
 Staten Island, New York 10314} 
\date{\today}

\begin{abstract}
Dark soliton (DS) and Josephson vortex (JV) in quasi-1D long Bose 
Josephson junction (BJJ) can be interconverted  
by tuning Josephson coupling. 
%This effect can reveal Josephson supercurrents even in 
%the overdamped regime. 
Rates of the interconversion as well as 
of the thermally activated phase-slip effect, resulting in the 
JV switching its vorticity, have been evaluated. The role of
quantum phase-slip in creating superposition of JVs with opposite
vorticities as a {\it qubit} is discussed as well.  
Utilization of the JV for controlled and coherent transfer of atomic 
Bose-Einstein condensate (BEC) is suggested. 

\end{abstract}


\maketitle

Solitons, in general, and the DS 
\cite{DS, snake, shlyap1}, in particular,
continue to be a fascinating 
subject for study.  
In 1D the snake instability \cite{snake}
is suppressed and the DS is a 
stable particle-like object (apart from the slow phonon induced decay 
\cite{shlyap1}).
Vortices in atomic BEC have been studied in great 
details theoretically and experimentally as well (see in ref.\cite{vortex}). 

Superfluid current circulation can exist in
two parallel quasi-1D waveguides, coupled by a 
uniform Josephson tunneling $\gamma>0$ along their length 
(see Fig.1), akin to the JV in superconducting long Josephson junction \cite{barone}.
Traditionally \cite{barone}, phase variation only is considered 
within the frame
of the Sine-Gordon (SG) equation. As it turns out, such description
is insufficient for  
the BEC waveguides in the {\it quasi}-1D regime, where 
the 1D Gross-Pitaevskii (GP) equation in the axial direction \cite{petrov}
should be employed.
Short BJJ have been thoroughly examined in refs.\cite{short,Leggett,Pitaevski}.
 Two coupled waveguides were already studied as well with focus on bright solitons \cite{mabop}. 
Quasi-1D BEC were created in a variety of 
magnetic traps \cite{hollow}. Intriguing perspectives
 are offered by BECs on microchips \cite{chip,twowi}. 
Coherence between two parallel elongated BECs has been demonstrated
in the seminal MIT experiment \cite{Kett1} and is currently being under
intense investigation \cite{Kett2,Kett3}.  
Two parallel waveguides with varying separation from each other were 
designed as an interferometer in refs.\cite{twowi,Kett2}. As noted in 
ref.\cite{mabop}, the considered 
setup has precise optical analogy -- dual-core optical fiber
 \cite{optdc}. 
Similar system with point-like coupling has been 
considered in ref.\cite{tworing} with emphasis on 
possible application in atomic interferometry. 

Here we show that as the ratio $\nu=\gamma /\mu$,
with $\mu$ being the chemical potential, 
becomes smaller than some critical
value $\nu_c$,
the DS transforms into the JV {\it spontaneously}. This breaks
the time-reversal symmetry. 
Conversely, as $\nu$ exceeds ~$\nu_c$, the JV transforms
into the DS, which restores the symmetry.   
The DS$\leftrightarrow$JV interconversion effect is a {\it reversible}
1D analog of the 3D DS snake instability \cite{snake}.
In contrast to the 3D, where the DS irrecoverably decays into
 vortex rings,
the DS in the quasi-1D BJJ can be controllably restored from the JV
by tuning $\nu$ above the critical value. The thermal and quantum 
phase-slip effects can restore the symmetry as well. 

\begin{figure}[h]
\begin{center}
\epsfxsize=8.5cm
\epsfbox{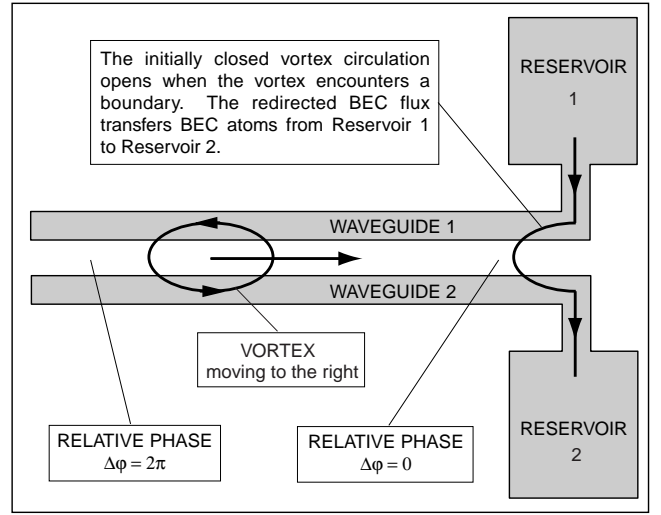}
\caption{
\label{transf}
Schematic representation of the long BJJ.
} 
\end{center}
\end{figure}

{\it Model: Action}. Dynamics of the bosonic fields 
~$\psi_k$, $k=1,2$, is controlled by the action
 \begin{equation}\label{S}
S=\int dt dx {\cal L},\quad {\cal L}={\cal L}_1 + {\cal L}_2 
+{\cal L}_{12},
\end{equation}
where $t$ is time and $x$ is the axial coordinate;
the Lagrangian density ~$\cal L$ consists of
the parts 
\begin{equation}\label{L_1}
{\cal L}_k=\Re [i\hbar\psi^*_k\dot{\psi}_k] 
-\frac{\hbar^2}{2m}|\nabla \psi_k|^2 + \mu_k|\psi_k|^2 
-\frac{g}{2}|\psi_k|^4,
\end{equation}
describing the dynamics along ~$x$,
as well as the Josephson tunneling
\begin{equation}\label{L_12}
{\cal L}_{12}=\gamma \psi^*_1\psi_2 + c.c.;
\end{equation}
chemical potentials $\mu_k=\mu$ in each waveguide will
be kept identical;
$m$ is atomic mass;
$g=4\pi\hbar^2 a/(m r^2_\perp)$--- 
the effective 1D interaction constant \cite{olshani},
with $a>0$ being 3D scattering length and $r_\perp$
standing for the effective width of the waveguides.
Strictly speaking, the standard Josephson coupling (\ref{L_12})
is valid for ~$\nu \ll 1$. As $\nu \sim 1$, higher
order terms in $\psi_{1,2}$ should be considered. However,
while such terms may change the
numerical value of ~$\nu_c$, they will not eliminate
the interconversion effect. Thus, in what follows,
we will not discuss the higher order couplings.

 {\it Model: Dissipative function}.
Significant damping of Josephson dynamics can occur
due to tunneling of the normal component \cite{Leggett}
between two BECs. Phenomenologically, the dissipation 
can be introduced through the dissipative
function ~$F_D$, so that
\begin{equation}\label{equation}
\frac{d}{dt}\frac{\delta \cal L}{\delta \dot{\psi}^*_k}-
\frac{\delta {\cal L}}{\delta \psi_k^*}
=-\frac{\delta F_D}{\delta \dot{\psi}^*_k}
\end{equation}
in accordance with the standard procedure \cite{Landau5}.
In general, $F_D$ must be positively defined function
of the time derivative of physically observable
quantities \cite{Landau5}. We choose it in the 
minimal form 
\begin{equation}\label{diss}
F_D=\int dx \frac{\dot{\rho}^2}{2\sigma}  ,
\end{equation}
where ~$\rho=|\psi_1|^2-|\psi_2|^2$.
All the information about normal component
is included into the kinetic coefficient ~$\sigma$, which
can be related
to the dissipation rate of small Josephson oscillations
considered in ref.\cite{Leggett}. Employing eqs.(\ref{S}-\ref{diss})
for uniform $\rho$ and small relative phase, we 
obtain damped Josephson oscillations
~$\ddot{\rho}+ \omega^2_J\rho + \kappa \dot{\rho}=0$, where
the Josephson frequency and the damping coefficient are 
~$\omega_J=2\sqrt{\gamma (\mu+2\gamma)}/\hbar$ and
~$\kappa=8\gamma (\mu+\gamma)/(\hbar^2 g\sigma)$, respectively.
The value of ~$\kappa$, which determines
a typical relaxation time ~$\tau \sim \sigma$,
 can be taken from the microscopic
analysis \cite{Leggett}.  

We introduce the units of length 
~$l_c=\hbar/\sqrt{m\mu}$ (correlation length), 
with $n_0=\mu/g$ being the average 1D density
in a single uncoupled ($\gamma=0$) waveguide, and of time ~$t_0=\hbar/\mu$.
Then, setting ~$\psi_k \to \sqrt{n_0}\psi_k $,
the action (\ref{S})
becomes $S=\hbar S_0 \int dt dx L$, where ~$S_0=l_c n_0$
(the validity of the GP regime is justified by
$S_0 \gg 1$ \cite{petrov});
~$L=L_1 + L_2 +L_{12} $, with
$L_k=i(\psi^*_k\dot{\psi}_k+ c.c.)/2  
-|\nabla \psi_k|^2/2 + |\psi_k|^2 
-|\psi_k|^4/2$ and 
$ L_{12}=\nu \psi^*_1\psi_2 + c.c.$.
The dissipative function density takes the form
 ~$\hbar S_0 \dot{\rho}^2/2\tilde{\sigma}$, with $\tilde{\sigma}=
\hbar \sigma t_0^2/n_0$.
Employing eqs.(\ref{S}-\ref{diss}) in these units,
we obtain 
\begin{eqnarray}
i\dot{\psi}_1 - \frac{1}{\tilde \sigma}\dot{\rho}\psi_1&=&
-\frac{\nabla^2}{2} \psi_1 - \psi_1 
+|\psi_1|^2\psi_1 - \nu \psi_2;
\label{eq_1}\\
i\dot{\psi}_2+ \frac{1}{\tilde \sigma}\dot{\rho}\psi_2
&=&-\frac{\nabla^2}{2} \psi_2 - \psi_2 
+|\psi_2|^2\psi_2 - \nu \psi_1.
\label{eq_2}
\end{eqnarray}
The dissipative terms ~$\sim \dot{\rho}$~ resemble 
the phenomenological dissipation introduced
in ref.\cite{Carlson}. These conserve the total number of atoms.
We note, however, that they {\it violate} the Galilean invariance 
(given by the transformation
~$\partial_t \to \partial_t -V\nabla_x$, $ \psi_k \to \exp[i(Vx+V^2t/2)]\psi_k$,
with $V$ being the velocity of a new frame moving along $x$).
In this regard, it is important to realize a limited nature of the
above phenomenological approach --- it can only be applied in the case
when the tunneling of the normal component {\it does not}
conserve linear momentum (along the waveguides), that is,
when scattering on imperfections of the trapping potential is
significant as in ref.\cite{Leggett}.   

{\it Dark soliton and Josephson vortex}.
Static one soliton solutions of eqs. (\ref{eq_1},\ref{eq_2})
 in infinite medium belong to a family
\begin{equation}\label{anz1}
\psi_{1,2}=\sqrt{1+\nu}\tanh(px)
\pm i\frac{B\sqrt{p/2}}{\cosh(px)}.
\end{equation}
The DS, characterized by ~$\psi_1=\psi_2$
 corresponds to $p=\sqrt{1+\nu}$ and $B=0$
(that is, total density has zero).
The static JV satisfies combined symmetry 
--- time reversal and reflection
($\psi_1=\psi ,\,\, \psi_2=\psi^*$), and, thus,
the equation
\begin{equation}\label{static}
(-\frac{1}{2}\nabla^2+|\psi|^2-1)\psi-\nu\psi^\ast = 0.
\end{equation}
Its solution (\ref{anz1}) 
is given by ~$p=2\sqrt{\nu}$ and $B\sqrt{p/2}=
\sqrt{1-3\nu}$. Obviously, it exists for ~$\nu<1/3$ only.
The phases ~$\varphi_{1,2}$ of the fields ~$\psi_{1,2}\sim \exp(i\varphi_{1,2})$
change from $\varphi_{1,2}=0$ at $x=-\infty$ to $\varphi_1=-\varphi_2=\pi$
at $x=+\infty$.

Eq.(\ref{static}) has been studied in a variety of completely different contexts and systems \cite{Ginz,Sarker,decay,GLsys,lcd,note1}. In terms of the real and 
imaginary parts, it was discussed in ref.\cite{Ginz} with respect to
the transformation of the
domain walls in magnetics. 
In the systems \cite{Ginz,Sarker,decay,GLsys}, the spontaneous transition 
between  the solutions of eq.(\ref{static}) has been found.
Its interpretation is system specific.

{\it DS-JV interconversion}. 
The DS formally exists for all values of the dimensionless
coupling $\nu$. The JV solution 
is valid only for $\nu<\nu_c=1/3$. At the critical value
$\nu_c$, the JV turns into the DS. Simple energy argument shows that the DS is an unstable state for $\nu<\nu_c$.
The energies $ E_{DS} = \frac{8}{3}(1+\nu)^{3/2}$
and $E_{JV} = \frac{8}{3}\sqrt{\nu}(3-\nu)$ of the DS and the JV,
respectively,  as well as their $\nu$-derivatives
become equal at $\nu=\nu_c$.
For $\nu <\nu_c$, one finds ~$E_{DS}>E_{JV}$, which
implies {\it absolute instability}
of the DS. 
 
Despite being identical to the problems of refs.\cite{Ginz,Sarker,decay,GLsys,lcd,note1}
in the static limit, dynamical eqs.(\ref{eq_1},\ref{eq_2}) cannot be mapped on these systems. Dynamics in our case is essentially two-component.
The interconversion can be well described
within the family (\ref{anz1}), where
$p=p(t),\,\, B=B(t)$ are some real and
complex functions of time, respectively.
Substituting (\ref{anz1}) into
eqs.(\ref{S}-\ref{diss}) and performing
the variational 
analysis based on the adiabatic
approximation with respect to the mass flow along the waveguides, 
we find for ~$\nu \to \nu_c$
\begin{eqnarray}
\dot{a}- \frac{16}{9}b -\frac{1}{\sqrt{3}}(a^2+b^2)b-
\frac{1}{\tilde{\tau}}\dot{b} =0,
\label{a}\\
\dot{b} + 3(\nu - \nu_c)a +\frac{1}{\sqrt{3}}(a^2+b^2)a=0, 
\label{b}
\end{eqnarray}
where $B=a+ib$, and $a,b$ are real, and 
~$ \tilde{\tau}= 3\tilde{\sigma}/(32(1+\nu))$.
It is quite obvious that the interconversion of the
DS ($a=b=0$) and the JV ($b=0,\,\, a=\pm 3^{3/4}\sqrt{\nu_c -\nu}$)
proceeds on typical relaxation time ~$\tau$~ of the BJJ \cite{Leggett}.
Very close to the instability (~$\nu \to \nu_c$)
the "critical" slowing down ~$\sim \tilde{\tau}/|\nu -\nu_c|$ takes place. 
Thus, the DS may vanish in accordance with the mechanism of ref.[3] 
before it decays into the JV. 
It is important to realize that the above result is independent
of a particular mechanism of dissipation.

\begin{figure}[h]
\begin{center}
\epsfxsize=8.5cm
\epsfbox{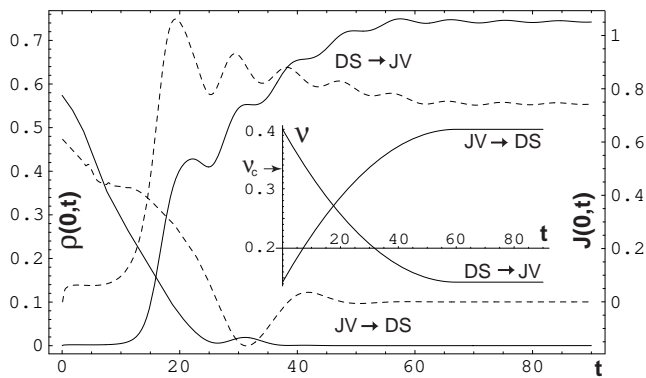}
\caption{
\label{phase}
Interconversion of the DS and JV is displayed via the evolutions of single 
waveguide density $\rho(x,t)$ (solid lines) and current 
$J(x,t)$ (dashed lines) at $x=0$. Slow change of $\nu$ passing through $\nu_c$ is shown on the inset. Transformation JV$\rightarrow$DS (DS$\rightarrow$JV) 
is manifested in the damped oscillatory transition 
of $\rho(0,t)$ and $J(0,t)$ from 
finite (zero) to zero (finite) values.
} 
\end{center}
\end{figure}

We have also performed direct 
numerical simulations of the full GP equations
(\ref{eq_1},\ref{eq_2}) with the initial conditions
taken as either DS or JV (located at $x=0$)
for periodic boundary
conditions, with the space period being about
10 times larger than soliton size. To accommodate
the phases variation by $\pi$, two-soliton
solutions were considered.
On Fig.\ref{phase}, the results of slow 
evolution of the coupling $\nu$ from below critical
$\nu=1/7$
(where JV is stable) toward above critical $\nu=2/5$ 
(where the DS is stable) as well as
its reverse is presented for the dissipation $\tilde{\sigma}=0.5$.
As perturbation, small uniform imbalance of
the waveguides population has been imposed on the initial
conditions.  
\noindent
Starting from JV and increasing $\nu$
causes damped phase-slip oscillations during
which the JV changes its vorticity (see curves marked
$JV \to DS$), and, finally, current vanishes. 
Correspondingly, the density, first, acquires zero at
finite time moments before the zero becomes permanent.
This final stage indicates formation of the DS.
If starting from the DS and decreasing $\nu$, the sign
of the JV vorticity is determined by sign of the initial
density imbalance. 
 We also note that the variational ansatz 
of 
%ref.\cite{our}
(\ref{a}, \ref{b})
 has been verified to reproduce the
full numerical solution with good accuracy for the initial imbalances
$\leq 10\%$.

{\it Vortex dynamics}. The moving DS solution is known analytically
(as gray
soliton ) \cite{DS}. Unfortunately, finding analytical
solution of the moving JV seems impossible. However, in the limit
~$\nu \to 0$, the JV can be well approximated by the SG
equation. Indeed, in this case, the variation of the
total density can be ignored. Thus, the representation
$\psi_{1,2}=\sqrt{1\pm \rho/2}{\rm e}^{\pm i\varphi/2}$
can be employed, with ~$|\rho| << 1$.
Substituting this into the (dimensionless) eqs.(\ref{S}
-\ref{diss}), and
ignoring the gradient $\nabla \rho$, we obtain after the variation
$
\dot{\varphi}- \rho - \tilde{\sigma}^{-1}\dot{\rho}=0$ and 
$
-\dot{\rho} + \nabla^2_x \varphi - 4 \nu \sin\varphi =0$.
Without dissipation ($\tilde{\sigma} \to \infty$),
we find $\rho=\dot{\varphi}$, and, then, 
the SG equation 
~$-\ddot{\varphi}+\nabla^2_x\varphi-4\nu\sin\varphi=0$ (see in \cite{barone}).
In the case $\tilde{\sigma} \to 0$, one finds
$\dot{\rho}=\tilde{\sigma}\dot{\varphi}$, and, then,
 the overdamped
SG equation \cite{barone}
~$-\tilde{\sigma}\dot{\varphi}+
\nabla^2_x\varphi-4\nu\sin\varphi=0$.
We note that, as ~$\nu \to 0$, the JV solution of eq.(\ref{anz1}) satisfies the static SG equation.

It is important to note that, in the SG approximation,
the JV is {\it always} stable. There is, simply, no room for the
DS due to the imposed constraints. 
Similar situation has been discussed
in ref.\cite{son} for two-component BEC. 
%There is, however, another
%limitation -- on speed $V$ of the JV. Indeed, before $V$
%approaches the sound barrier $V=1$, the energy of the 
%(underdamped) solution experiences "relativistic"
%growth ~$\sim \sqrt{\nu}/\sqrt{1-V^2}$. The size of the solution
%shrinks as ~$\sim \sqrt{1-V^2}/\sqrt{\nu}$. Thus, at some point,
%the size becomes comparable to the healing length $l_c$, and
%the phase slip will take place \cite{son}.

{\it Quasi-1D Fluctuations}. 
%A discussed in refs.\cite{petrov},
%he GP regime requires ~$S_0=l_cn_0 \gg 1$. 
At finite temperatures $T$,
phase-slip effects can destroy supercurrents. 
The corresponding life-time, however, can be very long \cite{svist1}.
In our case, stability of the JV is determined  
by the finite energy barrier $\Delta E=E_{DS}-E_{JV}$
with respect to thermal (quantum) jumps between the opposite orientations
of the current circulation. The probability of the thermal jump is
~$P\sim \exp(-\Delta E/T)$.  If ~$\nu \to \nu_c$, we find
\begin{equation}\label{P}  
P\sim \exp(-4.5\sqrt{3}(\nu_c - \nu)^2\frac{E^*}{T}).
\end{equation}
Thus, to have long lived JV, temperature must satisfy the
condition
\begin{equation}\label{fluc}  
T \ll 4.5\sqrt{3}(\nu_c - \nu)^2E^*,\quad \nu \to \nu_c.
\end{equation}
Here ~$E^*=\mu S_0=\sqrt{\mu T_c}$, with
~$T_c=\hbar^2n_0^2/m$~ being the temperature 
of the quasi-BEC \cite{svist2} formation.
For ~$\nu \to 0$, vortex-antivortex pairs can be created
thermally (quantum mechanically). To suppress the thermal effect,
~$T$~ must be less than the energy ~$2E_{JV}$ of the pair. Thus,
\begin{equation}\label{flu}  
T \ll 16\sqrt{\nu}E^*, \quad \nu \to 0. 
\end{equation}
It is important to note that, if the conditions (\ref{fluc},\ref{flu})
are met, the JV size ~$L_{JV}\approx l_c/\sqrt{\nu}$~ is smaller
than a typical phase coherence size ~$L_\phi=\hbar^2n_0/(mT)=l_cE^*/T$ of the
quasi-BEC \cite{svist2}.

At low temperatures,
quantum tunneling between two 
circulations of the JV will restore 
the symmetry for $\nu < \nu_c$ due to the
quantum phase-slip. This results in a
soliton, which, on one hand, is an 
essentially quantum object with respect
to its internal structure ---
a superposition of opposite vorticities, and,
on the other hand, can propagate like a heavy classical particle
($S_0 \gg 1$) along the BJJ-waveguides. It can be 
viewed as a mobile {\it qubit}.

{\it Vortex pump}. The vortex can be used to transfer a portion of the BEC atoms between BEC reservoirs (see Fig.\ref{transf}). The rate of the atom deposition/depletion in the first reservoir is ~$\dot{N}_1=J_1$, where ~$J_1=\rho_1\nabla_x \varphi_1$ is the current along the first waveguide taken at the point where
the coupling is zero (the right end of the waveguides). It is given by the density ~$\rho_1$ and the phase $\varphi_1$
at the boundary. Thus, $\Delta N= \int_{-\infty}^{+\infty} dt J_1$.
 If the reservoirs are large enough, the JV solution can be considered as
 undeformed at the boundary. Hence, one can set $J_1=J_1(x-x_0(t))$,
 where $x_0(t)$ is the position of the JV, and replace the integration
 ~$\int dt J_1=\int (dx/V(x))J_1$, with $V(x)=dx_0(t)/dt$. 
If the JV moves uniformly, this gives
%\begin{equation}
%\label{copu2}
%\Delta N =\frac{1}{ V}\int\rho_1\nabla\varphi_1 dx .
$\Delta N =\int\rho_1\nabla\varphi_1 dx /V$.
%\end{equation}
For small $V$, we substitute the static JV solution. Thus, the explicit integration
gives (in the physical units)
%\begin{equation}\label{DN}
$\Delta N =\frac{\pi S_0 V_s}{V}\sqrt{(1+\nu)(1-3\nu)}$,
%\end{equation}
where ~$V_s=\hbar /(m l_c)$ is the speed of sound. 
The GP regime ~$S_0 \gg 1$ \cite{petrov} implies ~$\Delta N \gg 1$.
%In reality, boundary with the reservoirs may modify eq.(\ref{DN}).

{\it Creation and detection of the JV}. 
Observation of the described interconversion can be based
on, first, creating the DS simultaneously
in the both waveguides. Then, moving them slowly apart (to have
~$\nu < \nu_c$) will result in 
vanishing of the DS into the JV, so that the zero in the densities will
heal (see Fig.\ref{phase}). Bringing the waveguides back together
(to have $\nu > \nu_c$)
will cause reappearing of the DS (see Fig.\ref{phase}).
The very fact of this reversible
transformation, besides being potentially utilized for creating 
the JV, can 
serve as an unambiguous evidence
of the static Josephson currents.
 It is important that
this effect, in contrast to the suggestion of ref.\cite{short},
 can be observed in the {\it overdamped} regime.
Direct imaging 
of the JV currents could be done by the Bragg spectroscopy 
technique \cite{bongs}. Analysis of the absorption imaging 
of the JV upon expansion
(for 3D vortices, see \cite{Kett4})
will be presented elsewhere. 

{\it Applications}. The utilizations of the JV
for
coherent BEC transfer and as a mobile qubit creates quite intriguing
perspectives for quantum computations and
coherent BEC manipulations in microtraps. 

{\it Acknowledgements}. 
We acknowledge useful discussions of the results with Boris Svistunov, Nikolay Prokof'ev and Eugene Chudnovsky. This work
was supported by the PSC-CUNY grant 64519-0033.


\begin{thebibliography}{99}

\bibitem{DS}
S. Burger, 
%K. Bongs, S. Dettmer, W. Ertmer, K. Sengstock, 
{\it et al}., Phys. Rev. Lett. {\bf 83}, 5198 (1999);
B.P. Anderson, {\it et al},
%P.C. Haljan, C.A. Regal, D.L. Feder, L.A. Collins, {\it et al}., 
{\it ibid}. {\bf 86}, 2926 (2001);
J. Denschlag, 
%J.E. Simsarian, D.L. Feder, C.W. Clark, L.A. Collins, 
{\it et al}., Science {\bf 287}, 97 (2000).

\bibitem{snake}
A.E. Muryshev, H.B. van Linden van den Heuvell, and G.V. Shlyapnikov, Phys. Rev. A {\bf 60}, R2665 (1999);
D.L. Feder, {\it et al},
%M.S. Pindzola, L.A. Collins, B.I. Schneider, and C.W. Clark, 
{\it ibid}. {\bf 62}, 053606 (2000);
Z. Dutton, {\it et al},
%M. Budde, C. Slowe, and L.V. Hau,
 Science {\bf 293}, 663 (2001).

\bibitem{shlyap1}
P.O. Fedichev, A.E. Muryshev, and G.V. Shlyapnikov, Phys. Rev. A {\bf 60}, 3220 (1999).

\bibitem{vortex}
A. Fetter and A. Svidzinsky, J. Phys.: Condens. Matter {\bf 13}, R135 (2001).

\bibitem{barone}
Barone, G.Paterno, {\it Physics and Applications
of the Josephson Effect}, John Wiley \& Sons, New York -
Singapore, 1982.

\bibitem{petrov}
D.S. Petrov, G.V. Shlyapnikov, and J.T.M. Walraven, Phys. Rev. Lett. {\bf 85}, 3745 (2000);
 {\it ibid}. {\bf 87}, 050404 (2001);
S. Dettmer, {\it et al},
%D. Hellweg, P. Ryytty, J.J. Arlt, W. Ertmer, {\it et al}.,
 {\it ibid}. {\bf 87}, 160406 (2001).

\bibitem{short}
A. Smerzi, {\it et al},
%S. Fantoni, S. Giovanazzi, and S. R. Shenoy,
Phys.Rev.Lett. {\bf 79}, 4950 (1997);
S. Giovanazzi, A. Smerzi, and S. Fantoni, {\it ibid}. {\bf 84}, 4521 (2000).
%S. Raghavan {\it et al}., Phys. Rev. A 59, 620 (1999).

\bibitem{Leggett}
I. Zapata, F. Sols and A.J. Leggett, Phys. Rev. A {\bf 57}, R28 (1998).

\bibitem{Pitaevski}
L. Pitaevskii and S. Stringari, Phys.Rev.Lett. {\bf 87}, 180402 (2000).

\bibitem{mabop}
V.S. Shchesnovich, B.A. Malomed and R.A. Kraenkel, 
Physica {\bf D188}, 213, (2004).

\bibitem{hollow}
K. Bongs, {\it et al},
%S. Burger, S. Dettmer, D. Hellweg, J. Arlt, {\it et al}.,
 Phys. Rev. A {\bf 63}, 031602(R) (2001);
A. Gorlitz, 
%J.M. Vogels, A.E. Leanhardt, C. Raman, T.L. Gustavson,
 {\it et al}., Phys. Rev. Lett. {\bf 87}, 130402 (2001).

\bibitem{chip}
For a review see
J. Reichel, Appl. Phys. B {\bf 74}, 469 (2002);
P. Treutlein, {\it et al},
%P. Hommelhoff, T. Steinmetz, T.W. Hänsch, J. Reichel,
 quant-ph/0311197.
%W. Hansel {\it et al}., Nature {\bf 413}, 498 (2001).
%A. E. Leanhardt {\it et al}., Phys. Rev. Lett. {\bf 90}, 100404 (2003);
%A. E. Leanhardt {\it et al}., {\it ibid}. {\bf 89}, 040401 (2002);

\bibitem{twowi}
H. Ott,{\it et al},
% J. Fortagh, G. Schlotterbeck, A. Grossmann, and C. Zimmermann,
 Phys. Rev. Lett. {\bf 87}, 230401 (2001);
E.A. Hinds, C.J. Vale, and M.G. Boshier, {\it ibid}. {\bf 86}, 1462 (2001);
For a review see
L. Feenstra, L.M. Andersson, J. Schmiedmayer, cond-mat/0302059 (2003).

\bibitem{Kett1}
M.R. Andrews, {\it et al}, 
%C.G. Townsend, H.-J. Miesner,
% D.S. Durfee, D.M. Kurn, and W. Ketterle,
Science {\bf 275}, 637-641 (1997).

\bibitem{Kett2}
Y. Shin, {\it et al}, 
%M. Saba, T.A. Pasquini, W. Ketterle, D.E. Pritchard, A.E. Leanhardt, 
Phys. Rev. Lett. {\bf 92}, 050405 (2004).

\bibitem{Kett3}
Y. Shin, 
%M. Saba, A. Schirotzek, T. A. Pasquini, A. E. Leanhardt,
 {\it et al}, Phys. Rev. Lett. {\bf 92}, 150401 (2004).


\bibitem{optdc}
M. Romagnoli, S. Trillo, and S. Wabnitz, Opt. Quantum Electron. {\bf 24}, S1237 (1992);
A. Ankiewicz, M. Karlsson and N. Akhmediev, Optics Communications {\bf 111}, 116 (1994).

\bibitem{tworing}
B.P. Anderson, K. Dholakia, and E.M. Wright, Phys. Rev. A {\bf 67}, 033601 (2003).

\bibitem{olshani}
M. Olshanii, Phys. Rev. Lett. {\bf 81}, 938 (1998).

\bibitem{Landau5}
L.D. Landau, E.M. Lifshitz, L.P. Pitaevskii, {\it "Statistical Physics"},
 Butterworth-Heinemann 3rd edition (1999), p.368.

\bibitem{Carlson}
N.N. Carlson, Physica {\bf D98}, 183 (1996).

\bibitem{Ginz}
L. N. Bulaevski and V. L. Ginzburg, Sov. Phys. JEPT {\bf 18}, 530 (1964)]
\bibitem{Sarker}
S. Sarker, S.E. Trullinger and A.R. Bishop, Phys. Lett. A {\bf 59}, 255 (1976).

\bibitem{decay}
P. Coullet and J. Lega, Phys. Rev. Lett. {\bf 65}, 1352 (1990).
%J. Lajzerowicz and J.J. Niez, J. de Phys. {\bf 40}, L165 (1979)

\bibitem{GLsys}
L. Korzinov, M.I. Rabinovich, and L.S. Tsimring , Phys. Rev. A {\bf 46}, 7601 (1992);
T. Frisch,{\it et al}, 
%S. Rica, P. Coullet, and J.M. Gilli, 
Phys. Rev. Lett. {\bf 72}, 1471 (1994).

\bibitem{lcd}
B. Denardo, W. Wright, and S. Putterman, Phys. Rev. Lett. {\bf 64}, 1518 (1990);
B. Denardo,{\it et al}, 
%B. Galvin, A. Greenfield, A. Larraza, S. Putterman,
% and W. Wright, 
{\it ibid}. {\bf 68}, 1730 (1992);


\bibitem{note1}
I.V. Barashenkov, S. R. Woodford, and E. V. Zemlyanaya,
Phys. Rev. Lett. {\bf 90}, 054103 (2003)

%\bibitem{our}
%V.Kaurov and A.B.Kuklov, cond-mat 0312084.
\bibitem{son}
D.T. Son and M.A. Stephanov, Phys. Rev. A, {\bf 65}, 063621 (2002).

\bibitem{svist1}
Yu. Kagan, N.V. Prokof'ev, and B.V. Svistunov, Phys.Rev. A, {\bf 61}, 045601(2000).

\bibitem{svist2}
Yu.\ Kagan, B.V.\ Svistunov, and G.V.\ Shlyapnikov, 
Sov. Phys. - JETP {\bf 66}, 314 (1987);
D.S.\ Fisher and P.C.\ Hohenberg, 
Phys. Rev. B {\bf 37}, 4936 (1988).

\bibitem{bongs}
K. Bongs, {\it et al}
%S. Burger, D. Hellweg, M. Kottke, S. Dettmer, {\it et al}., 
J.Opt. B {\bf 5}, S124 (2003).

\bibitem{Kett4}
S. Inouye, 
{\it et al}
%S. Gupta, T. Rosenband, A.P. Chikkatur, A. Görlitz, {\it et al}.,
%Observation of vortex phase singularities in Bose-Einstein condensates.
Phys. Rev. Lett. {\bf 87}, 080402 (2001).

\end{thebibliography}
\end{document}